\def\mytitle{Dual-Quaternion Julia Fractals}
\def\mysubtitle{}
\def\myauthor  {Ben Kenwright}

\def\mykeywords{graphics, dual-quaternion, visualization, fractals, geometry, graphical processing unit}

\documentclass[tog,backref=page]{acmsiggraph}
\TOGonlineid{45678}
\TOGvolume{0}
\TOGnumber{0}
\TOGarticleDOI{1111111.2222222}
\TOGprojectURL{}
\TOGvideoURL{}
\TOGdataURL{}
\TOGcodeURL{}

\usepackage[compact]{titlesec}
\titlespacing{\section}{0pt}{0ex}{0ex}
\titlespacing{\subsection}{0pt}{0ex}{0ex}
\titlespacing{\subsubsection}{0pt}{0.0ex}{0ex}

\usepackage{listings}

\usepackage{amsmath} 

\usepackage{subcaption}

\usepackage{enumitem}
\setlist[itemize]{noitemsep,leftmargin=*}
\setlist[enumerate]{noitemsep,leftmargin=*}

\usepackage{caption}
\usepackage{subcaption}

\usepackage{listings}

\hypersetup{
	colorlinks  = true, 
	linkcolor   = blue,
	anchorcolor = red,
	citecolor   = blue, 
	filecolor   = red, 
}

\usepackage{graphicx}

\graphicspath{{./images/}{./}}

\newcommand{\figuremacroW}[4]{
	\begin{figure}[h] 
		\centering
		\includegraphics[width=#4\columnwidth]{#1}
		\caption[#2]{\textbf{#2} - #3}
		\label{fig:#1}
	\end{figure}
}

\newcommand{\figuremacroF}[4]{
	\begin{figure*}[t] 
		\centering
		\includegraphics[width=#4\textwidth]{#1}
		\caption[#2]{\textbf{#2} - #3}
		\label{fig:#1}
	\end{figure*}
}

\usepackage{amssymb}

\usepackage[ruled]{algorithm2e}

\usepackage{amsmath} 

\usepackage{setspace}

\title{\mytitle\\ \fontsize{12pt}{22pt}\selectfont \vspace{-5pt} \mysubtitle}

\author{\myauthor\\
}

\pdfauthor{}

\teaser{
\centering
    \includegraphics[width=1.0\textwidth]{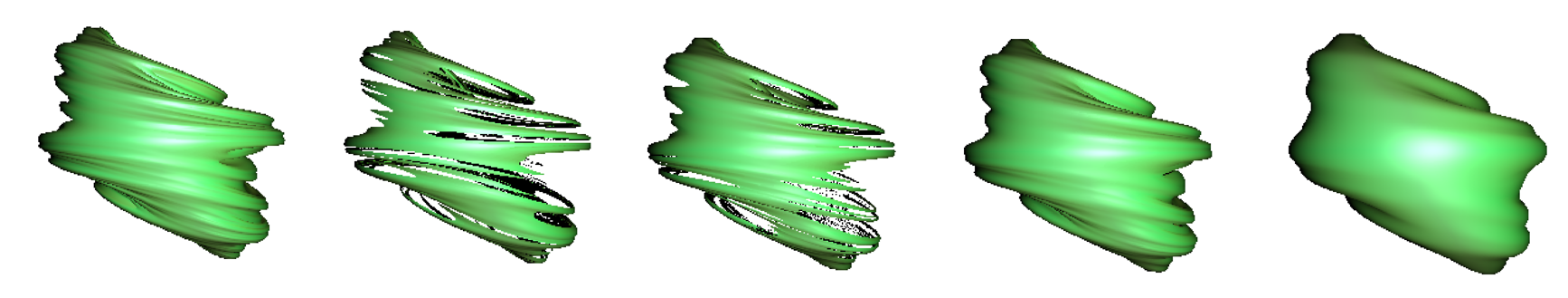}
     \caption{Exploratory parameter analysis in early prototype Julia sets of quadratic fractals with Dual-Quaternions.}
     \label{fig:teaser}
}

\begin{document}
	
\raggedbottom
\sloppy

\widowpenalties 1 0
\raggedbottom
\sloppy

\maketitle

\begin{abstract}

Fractals offer the ability to generate fascinating geometric shapes with all sorts of unique characteristics (for instance, fractal geometry provides a basis for modelling infinite detail found in nature). 
While fractals are non-euclidean mathematical objects which possess an assortment of properties (e.g., attractivity and symmetry), they are also able to be scaled down, rotated, skewed and replicated in embedded contexts. 
Hence, many different types of fractals have come into limelight since their origin discovery. 
One particularly popular method for generating fractal geometry is using Julia sets.
Julia sets provide a straightforward and innovative method for generating fractal geometry using an iterative computational modelling algorithm.
In this paper, we present a method that combines Julia sets with dual-quaternion algebra. 
Dual-quaternions are an alluring principal with a whole range interesting mathematical possibilities.
Extending fractal Julia sets to encompass dual-quaternions algebra provides us with a novel visualize solution.
%
%
%
%
We explain the method of fractals using the dual-quaternions in combination with Julia sets.
Our prototype implementation demonstrate an efficient methods for rendering fractal geometry using dual-quaternion Julia sets based upon an uncomplicated ray tracing algorithm.
We show a number of different experimental isosurface examples to demonstrate the viability of our approach.

\end{abstract}

\keywordlist

\setlength{\ACMcopyrightspace}{1.1in}


\section{Introduction}

%

\paragraph{Geometric Properties of Complex \& Dual Numbers}
Complex numbers have a very elegant geometric interpretation. Specifically, we can treat complex numbers as vectors in the complex plane.  Addition and subtraction then follow regular rules of vector arithmetic, while complex multiplication can be seen as scaling the rotation of one vector by the magnitude and argument of another, respectively. A dual number is a number of the form $a+b\varepsilon$, where $a, b \in \mathbb{R}$ and $\varepsilon$ is a non-real number with the property $\varepsilon^2=0$.  Dual numbers are in some way similar to complex numbers $a+bi$, where $i^2=-1$. 
You can imagine dual numbers as situated on the orthogonal line centred at `a' in the dual number plane.  When treating dual numbers as vectors the modulus is evaluated as $||a+b\varepsilon||  = abs(a)$ because of the nilpotency \cite{wang2010feedback}.

\paragraph{Dual-Quaternions}
Dual-quaternion have been around since 1882 \cite{clifford1882mathematical} but has gained less attention compared to quaternions alone. 
Comparable to quaternions the dual-quaternions have had a taboo associated with them, whereby individuals avoid quaternion and hence dual-quaternions \cite{kenwright2012beginners}. 
While the robotics community has started to adopt dual-quaternions in recent years, the computer graphics community has not embraced them as whole-heartedly. 
Recent work which has taken hold and has demonstrated the practicality of dual-quaternions in computer graphics, some examples include:
Kavan et al. \cite{kavan2008geometric} who demonstrated the advantages of dual-quaternions in character skinning and blending.
Frey and Herzeg \cite{frey2011spherical} extended Kavan et al.'s \cite{kavan2008geometric} work with dual-quaternions and qtangents as an alternative method for representing rigid transforms instead of matrices. Giving evidence that the results can be faster with accumulated transformations of joints if the inferences per vertex are large enough.
Selig \cite{selig2010rational} address key limitations in computer games by examining the problem of solving the equations of motion in real-time by putting forward how dual-quaternion give a very neat and succinct way for representing rigid-body transformations.
Vasilakis and Fudos \cite{vasilakis2009skeleton} discussed skeleton-based rigid-skinning for character animation.
Kuang et al. \cite{kuang2011strategy} presented a strategy for creating real-time animation of clothed body movement.

\paragraph{Algebra and Fractals}

Julia set of a function $f$ is the set of all points $z$ in $C$ such that $f$ displays sensitive dependence at $z$.
A popular Julia set function, and the one we use in this paper, is of the form $f(z) = z^2 + c$ for some constant $c$.
In the two dimensional complex domain, a Julia set is produced using the iterative algorithm $z_{n+1} = z_n^2 + c$, for a given complex number $c$. 
The Julia set contains the initial complex values $z_0$ for which the system converges. 
There have been a number of publications devoted to computational and graphical aspects of the Julia sets \cite{ke1990journey,crane2005ray}. 
including 3-dimensional and 4-dimensional contexts to generate fractal geometry.
In particular, modelling fractal geometries using quaternions is a well researched topic in computer graphics \cite{pickover1998chaos}.
Along similar lines, Octonionic Julia sets have also been investigated in the literature from the computational and graphical point of view.
Dual-quaternions follow a regular set of arithmetic rules (for performing common operations, such as, multiplication, addition and magnitude - see Appendix \ref{sec:appendix} and Figure \ref{fig:dual}).
Extending the concepts developed for creating quaternion Julia sets, allows us to formulate similar methods for dual-quaternion Julia sets.

\paragraph{Contribution}
The key contributions of this paper are:
(1) process for building dual-quaternion Julia sets;
(2) graphical examples and experiments.

\figuremacroW
{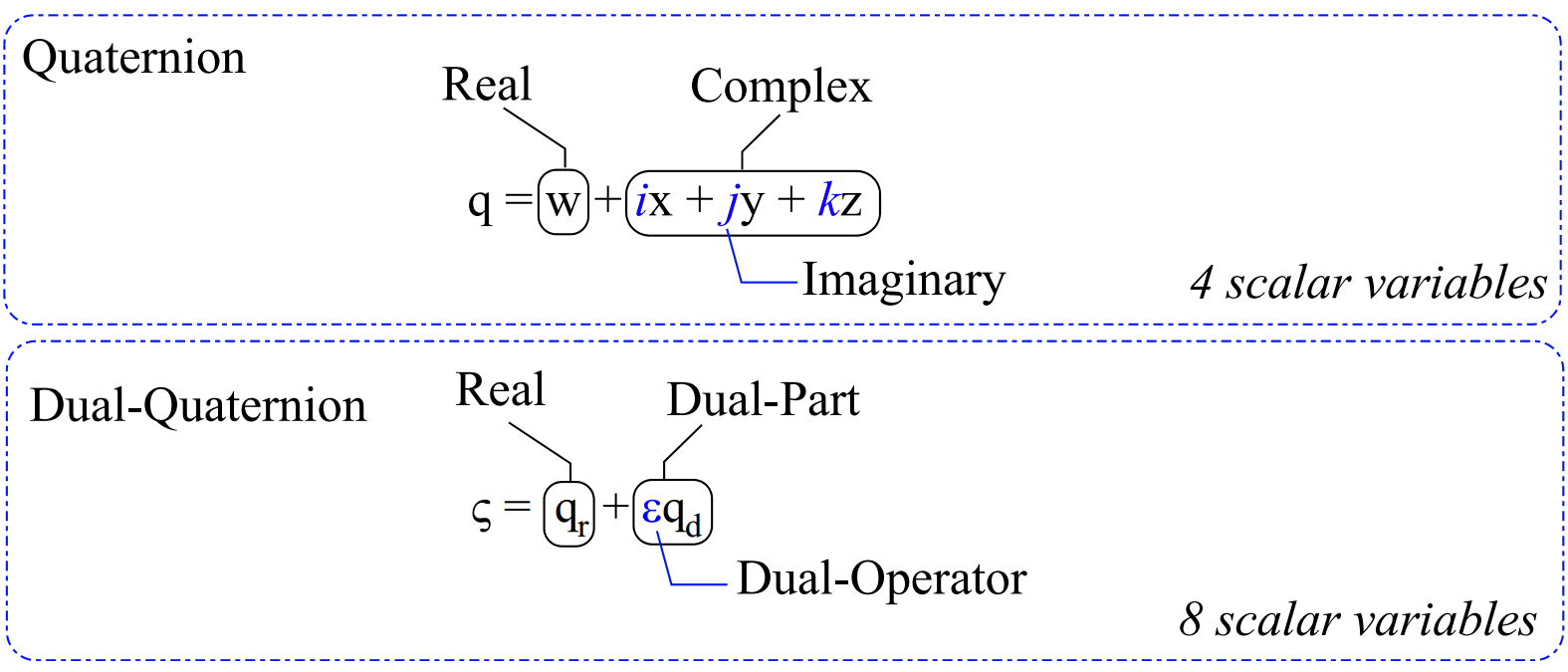}
{Dual-Quaternion}
{Visual Overview of Quaternion and Dual-Quaternion Components.  To avoid confusion and enable the reader to easily distinguish a quaternion from a dual-quaternion we use two discernible symbols to identify them. }
{1.0}

\section{Related Work}


\paragraph{Quaternions to Dual-Quaternions}

Clifford \cite{clifford1882mathematical} published his intriguing work on dual-numbers
in 1873, and provided us with a powerful tool for
facilitating the analysis of complex systems (e.g.,
mechanical, geometric) \cite{kenwright2012beginners}.
The dual-quaternion is an extension of dual-number theory
whereby the numbers for the dual-number equation are
represented by quaternions. Remarkably, the dual-quaternion algebra that results is very straightforward and
elegant and provides us an algebraically compact and efficient system for solving otherwise complex problems \cite{kenwright2012beginners}.
Dual numbers are pretty similar to imaginary numbers but there is one important difference. With imaginary numbers, $i^2 = -1$, but with dual numbers, $\varepsilon^2 = 0$ (and $\varepsilon$ is not 0), while this may seem like a small difference, it opens up a whole interesting world of mathematical possibilities.

\paragraph{Fractal Geometry}

Generating a 4-dimensional version of complex numbers are called quaternions. 
Alan Norton \cite{norton1982generation} was the first to demonstrate the application of the Quaternion Julia sets made by displaying a 3D `slice' of the 4D space. 
%
Quaternion Julet sets is actually a projection from 4 dimensions to 3 dimensions, akin to how a 2-dimensional square is a presentation of a 3-dimensional projection of a cube. 
Despite the added complexity, it possesses an underlying smooth appearance (i.e., less interesting fractal detail when you zoom in compared to the 2-dimensional version).
As we show in this paper higher-dimensional maths can be used to create 3-dimensional Juliet set fractals. 
Daniel White's `Mandelbulb' \cite{online2007} takes a different approach. He took the geometrical properties of the `complex plane' where multiplication becomes rotation and addition becomes movement of the plane in a particular direction and applied them to a 3- dimensional space.
The concept allows the generation of striking visuals of how an apparently simple process can lead to highly intricate sets. 
There are numbers of preceding studies about properties of the fractals and Julia sets and how to implement them in computer graphics \cite{crane2005ray}. 
Similar to the existing results we ray-trace the interpolation Julia set for making shading-reflecting like effect and apply the proposed method into the stereographic projection of complex space to obtain a kind of uneven and shaded surfaces. 


\figuremacroW
{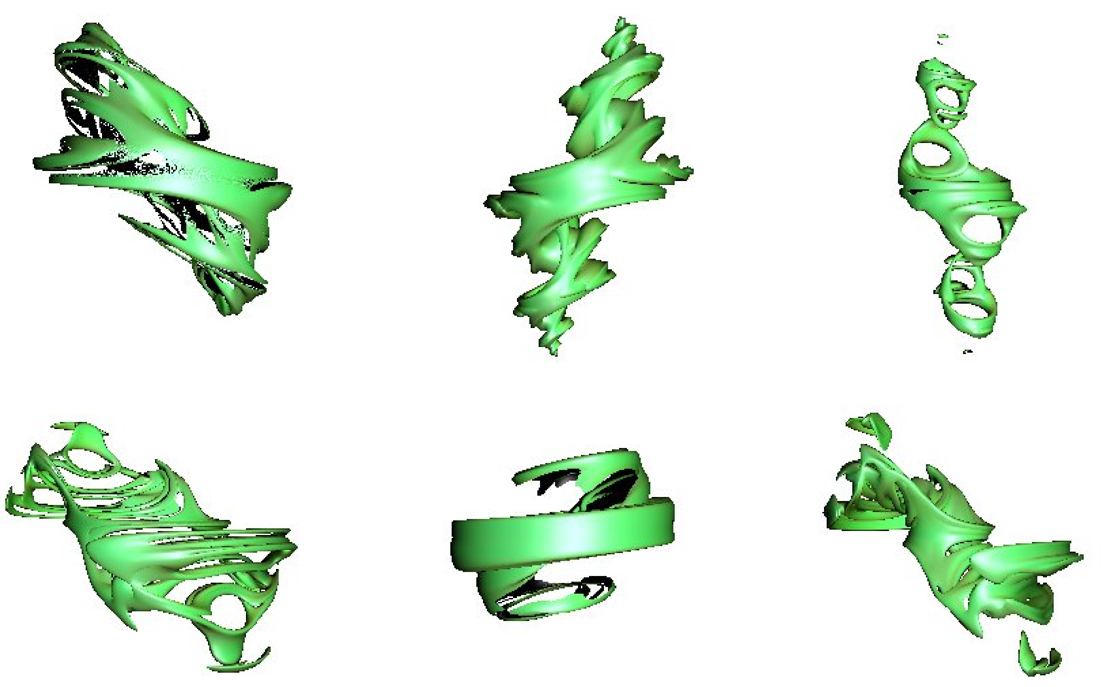}
{Experimental}
{Exploring visual possibilities through random parameters.
Parameter constant value from top left to bottom right:
c=(-0.10,0.8,-0.26,0.15)(-023,-0.38,-0.86,0.64);
c=(-0.67,-0.54,-0.07,0.02)(0.06,-0.53,0.15,-0.27);
c=(-0.98,0.27,0.40,0.20)(-0.37,-0.41,-0.24,-0.34);
c=(0.35,0.78,0.85,-0.57)(-0.22,0.06,-0.46,0.05);
c=(-0.17,-0.28,0.11,0.8)(0.06,0.44,-0.66,0.06);
c=(-0.04,0.95,0.4,-0.43)(0.09,-0.45,-0.27,-0.31).
}
{1.0}

\section{Method}

\paragraph{Julia Set Fractals}

Gaston Julia was the French mathematician who made the exciting discovery of Julia Sets in 1918 \cite{mandelbrot1983fractal}.
Julia sets are a popular fractal formed using formula iterations.
The Julia set is the union of all points $z$ in an iterated complex function $f(z)$ that do not diverge as the number of iterations of the function approach infinity (set of points form a `filled' in volume).
Julia sets forms a boundary using the the filled-in set.
There are a large range of functions that encompass Julia sets, however, in this paper, we employ the popular quadratic expression given in Equation \ref{eq:julia}

\begin{equation}
	f(z) = z_{n+1} = z_{n}^{2}+c
\label{eq:julia}
\end{equation}

\noindent where $z$ represents a variable of the form $a+bi$ ($a$ and $b$ are real-numbers with $i$ an imaginary number).
The value $c$ is also a complex number and is constant for each unique Julia.
Hence, as each value of $c$ gives a different Julia set there are an infinite number of sets.
In an uncomplicated case, the value $a+bi$ represents a 2-dimensional complex plane with each starting point $z_0$ for the iterative series representing a pixel position.
For each starting value there are two possibilities for what will happen, either (1) as $n$ increases $f(z)$ will tend towards infinity or (2) the value will stay within a bounded value.
Points which do not stay within the bounding limit are said to be in the escape set, while all other points are termed prisoners (prisoner set).
The common boundary between the escape set and the prisoner set is called the Julia set (defined for a particular $c$).

\paragraph{Quaternion Julia Sets}

A quaternion is represented by two fundamental parts, a
scalar real part ($w$) and an imaginary vector part ($x,y,z$).
%
We can extend the recursive Julia set algorithm given in Equation \ref{eq:julia} to use quaternions as given in Equation \ref{eq:dqjulia}.  
However, we need the squared quaternion, which is given in Equation \ref{eq:quatsq}.
The square of a quaternion is taken from previous quaternion geometric research \cite{hart1989ray,hart1990interactive} with Julia sets (takes advantage of the trigonometric properties of quaternions).

\begin{equation}
\begin{alignedat}{3}
q   &= (\vec{v}, w) = (x, y, z, w) \\
q^2 &= (x^2-y^2 - z^2 - w^2, \; 2xy, \; 2xz, \; 2xw ) 
\end{alignedat}
\label{eq:quatsq}
\end{equation}

The forward Julia iteration is given by Equation \ref{eq:dqjulia}:

\begin{equation}
	q_{n+1} = q_n^2 + c
\label{eq:dqjulia}
\end{equation}

\figuremacroF
{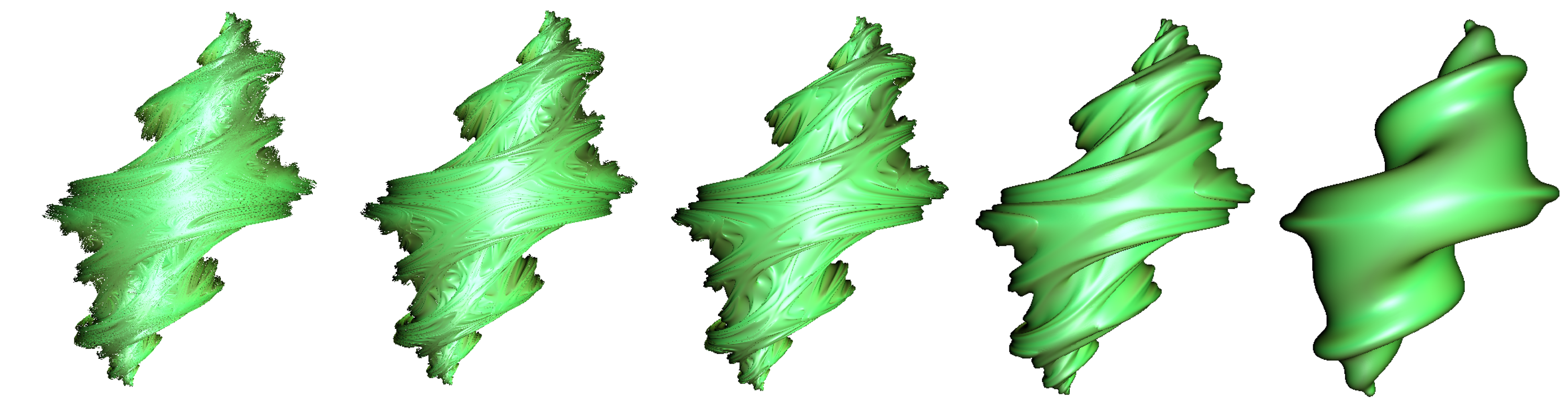}
{Level of Detail}
{Trade-off between additional detail and computational cost - number of sample points and maximum iterations (n) for determining if a point is inside or outside the Julia set.}
{1.0}

\paragraph{Dual-Quaternion Julia Sets}

Dual-quaternion Julia sets is the same as ordinary Julia sets except that dual-quaternions are used instead of complex numbers.
Generation of the fractal , the function $\zeta_{n+1}=f(\zeta)$ is iterated and if the series is bounded when $n$ tends to infinity then the dual-quaternion $\zeta_0$ belongs to the Julia set.

\begin{equation}
\begin{alignedat}{3}
\zeta   &=  q_a+q_b\varepsilon \\
\zeta^2 &= {q_a^2+q_b^2\varepsilon} 
\end{alignedat}
\end{equation}

\noindent where $a$ and $b$ are the dual-quaternion parameters (i.e., real and dual number represented as two quaternions).  We apply Equation \ref{eq:quatsq} to calculate the `squaring' of each component and add the dual-quaternion constant to achieve each forward iteration given by the Julia Equation \ref{eq:dqjulia2}.

\begin{equation}
\zeta_{n+1} = \zeta_n^2 + c
\label{eq:dqjulia2}
\end{equation}

\noindent where $\zeta$ and $c$ are dual-quaternions; $n$ is the number of iterations, typically between $6$ and $15$ (Figure \ref{fig:teaser} shows the impact of modifying the number of iterations on the level of detail).

\paragraph{Practical Considerations}

Impossible to iterate each sample point to infinity - hence, a suitable maximum iteration range is defined.
If the series has not reached the escape limit by then, the point is considered inside the set.
In practice, a fairly small number of iterations gives reasonably good results \cite{hart1989ray} (e.g., less than 10).
However, the more iterations the more accurate the fractal surface.
As shown in Figure \ref{fig:detail}, the more iterations and the finer the resolution provide additional detail for the geometric model at increased computational cost.

\paragraph{Dimension Reduction}
Dual-quaternion Julia sets are in 8-dimensional space.
How to visualize the dual-quaternion Julia set in 3-dimensional space.
We need to map the 8-dimensions into 3-dimensional space.
There are two approaches:
(1) render a single 3-dimensional slice of the dual-quaternion space or
(2) project the dual-quaternion space onto 3-dimensions to see a 3-dimensional shadow of the Julia set.
The approach used in this paper is to intersect the dual-quaternion space with a plane; essentially making five of the dual-quaternion components (dimensions) constant values.

\paragraph{Ray-Marching}

Reduce the computational cost of testing every point in the 3-dimensional space.
Slow to generate a model of acceptable granularity in a reasonable space of time.
For example, see comparisons with a Voxel-based render shown in Figure \ref{fig:voxels}. 
To offer a fast and efficient visualization solution, we utilize a ray-marching algorithm.
The ray-marching algorithm is able to produce high-quality detailed renders of the dual-quaternion Julia set at various levels.
We employ the distance estimation method for the Julia set \cite{dang2002hypercomplex}.
This distance estimator can be used to accelerate the ray tracing process using unbounding volumes.
The estimation gives a lower-bound on the distance to the fractal from any given point outside the fractal.
While the Equation \ref{eq:distfract} provided an initial approximation, we found the calculation from Hart et al. \cite{hart1989ray} given in Equation \ref{eq:distfract2} provided more aesthetically pleasing results (and required less tweaking/tuning). 

\begin{equation}
\begin{alignedat}{3}
d \ge \alpha \left( \frac{|\zeta_n|}{|\zeta_n'|} \right) \\
|\zeta_n'| = 2 |\zeta_{n+1}| |\zeta_{n+1}'|
\end{alignedat}
\label{eq:distfract}
\end{equation}

\noindent where $\alpha$ is analogous to the grid size and typically has a small value (e.g., $\le 0.1$).
Once the lower-bound on the distance is known, we are able to step along the ray until we hit the fractal or have stepped to many times without hitting the fractal.
An additional optimization is to have rays intersect with a sphere encapsulating the fractal first.
This helps speed up the ray tracing algorithm to quickly discard rays that would not intersect the fractal.

\begin{equation}
\begin{alignedat}{3}
d \ge  0.5 \frac{ |\zeta_n| } {|\zeta_n'|} log( |\zeta_n| ) 
\end{alignedat}
\label{eq:distfract2}
\end{equation}

\figuremacroW
{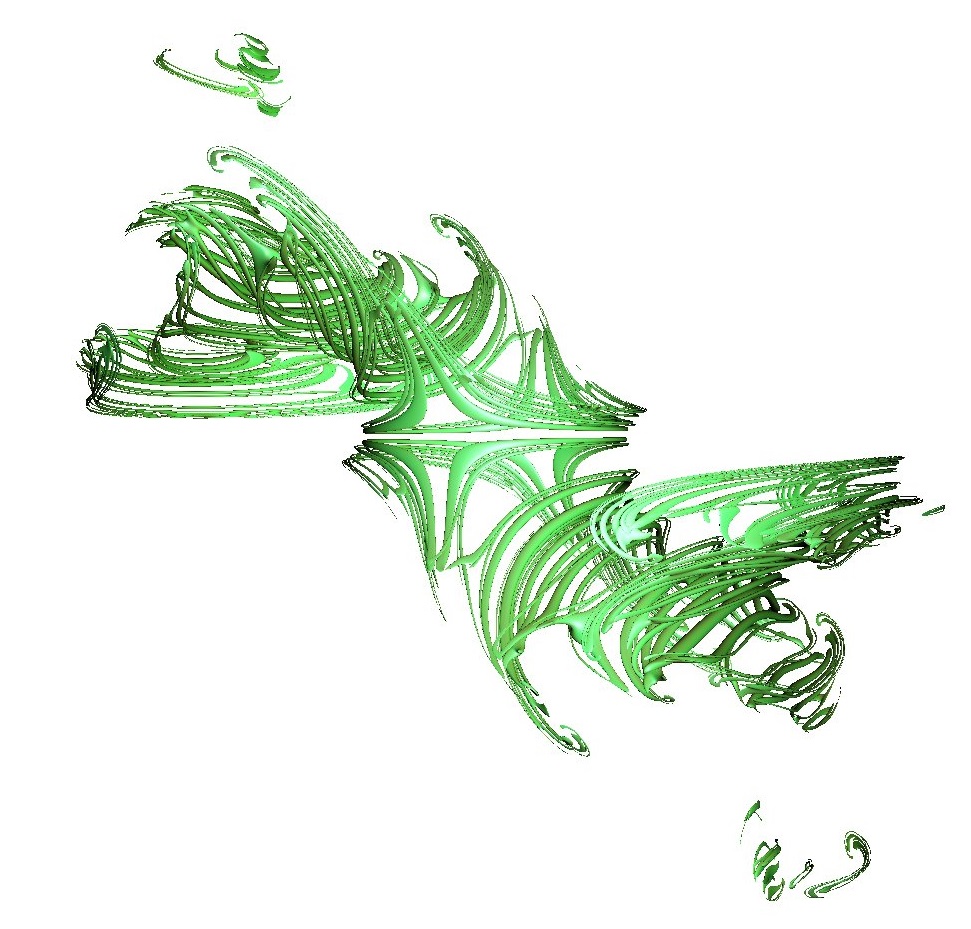}
{High-Detail}
{Emphasising detail within the dual-quaternion Julia set geometry by using a larger number of iterations. c =  (-0.04,0.95,0.4,-0.43)(0.09,-0.35,-0.27,-0.31), n=15.}
{1.0}

\paragraph{Lighting/Illumination}

Shading is an extremely important process in creating a realistic and aesthetically pleasing image. 
%
As we are rendering a 3-dimensional model, lighting is crucial to visually identify differences in geometric detail.
We employ the Phong model as it is straightforward and easy to incorporate into the ray-tracing algorithm.
However, the Phong model requires the surface normals.
We calculate a rough surface normal by sampling the surrounding points (i.e., above, below, left and right).
When a point in the Julia set is found the four neighbouring points are also tested to see if they are inside or outside the set.
These points are required for generating the neighbouring surface that allows us to compute the surface normal ($\vec{n}$).
We calculate the divergence at each pixel by iterating several points with a small delta.
While the slope at any point on the surface of the Julia set is undefined, closer examination of a points surroundings, for example scanned at higher iteration and z-resolution, would reveal a different environment. An exact normal can not be calculated, however the approximation using a sample set of the surrounding points provides an acceptable surface normal for our experiments.
%
The Phone illumination calculation for a single light is given below in Equation \ref{eq:phong}.

\begin{equation}
I = k_a I_{a} + k_d I_{d} (\vec{n} \cdot \vec{l}) + k_s (\vec{r} \cdot \vec{v})^{s}
\label{eq:phong}
\end{equation}

\noindent where
$I$ is the total illumination,
$k$ is the attenuation coefficient for the object material (ambient, specular and diffuse),
$\vec{l}$ is the vector to the light,
$\vec{r}$ is the reflected light vector (i.e., $R = 2 (\vec{n} (\vec{n} \cdot \vec{l})) - \vec{l}$ ),
$\vec{n}$ is the normal vector at the object surface,
$\vec{v}$ is the vector from the eye point,
$s$ is the specular exponent,
and
$I_a, I_s, I_d$ are the light intensities.

\figuremacroW
{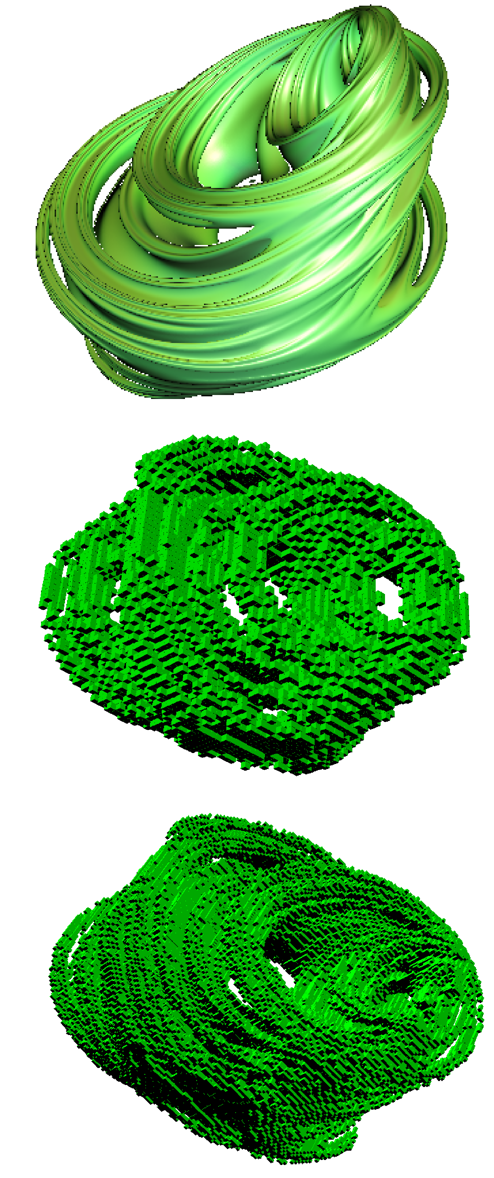}
{Voxel vs Ray-Traced}
{Initial prototypes tried to utilize Voxels on the GPU \cite{zadick2016integrating} (i.e., geometry shader) - to create a real-time prototyping tool for experimentation.  However, a ray-tracing algorithm was more suitable for emphasising detail due to computational/resolution limits.  Top ray-traced image, middle $50 \times 50 \times 50$ voxel grid and bottom $100 \times 100 \times 100$ voxel grid.}
{0.7}

\paragraph{Graphical Processing Unit (GPU)}

Fractal rendering has a high computational costs and can be parallelised easily. 
For example, Keenan \cite{crane2005ray} demonstrated the generation of quaternion Julia sets using a GPU for the ray-tracing algorithm (fragment shader). 
We employ a similar approach with the ray-marching algorithm on the GPU used to render the geometric surface of the dual-quaternion Julia set.


\section{Experimental Results}

\paragraph{Parameter Tuning/Experimenting}

Figure \ref{fig:teaser} and Figure \ref{fig:tests} show some preliminary experimental results through parameter tuning/changing.
As shown in Figure \ref{fig:detail}, adjusting the maximum number of iterations and the in/out Julia set range allows us to capture additional low-level details (i.e., at additional computational cost and the moire effect due to the problem of too much detail in the image/model).
%
%
The simulations provide a strong correlation in style and appearance to quaternion Julia sets (twirling candy floss appearance).
While the dual-quaternion 
space has a clearly defined sum, product and norm -  the distance squared estimator in this paper uses a quaternion calculation for the real and dual components which means it still has a strong coupling to a 4-dimensional space. 
Of course, this leaves the door open for further exploratory work into other distance square estimator variations using dual-quaternion algebra.

\section{Conclusion/Discussion}

In conclusion, fractals are a powerful tool for generating interesting patterns.
Extending the Julia set to higher dimensions allows us to form shapes with strong stereoscopic properties which exhibit intertwisting, irregular fringes and transformations.
In this paper, we employed dual-quaternion algebra to realize a 3-dimensional fractal using the Julia set.
When trying to capture the dimensional properties of dual-quaternions using Julia sets, we rendered snapshots from different sample points. 
There is still lot to be discovered about Julia sets in the dual-quaternion space.
The rules defined by dual-quaternion algebra are applicable to Julia set to extend the concept to multiple dimensions. 
Techniques for visualizing the higher dimensional Julia set as a three dimensional object. 
We describe our attempt to extend the already established technique of ray tracing Julia sets to incorporate dual-quaternion properties. 
We also discussed optimization algorithm for reducing computational times through exploitation of the graphical processing unit (allowing for exploratory prototyping). 
The scientific value of these images can be questioned, but they possess aesthetic factors of fascination. 

\figuremacroW
{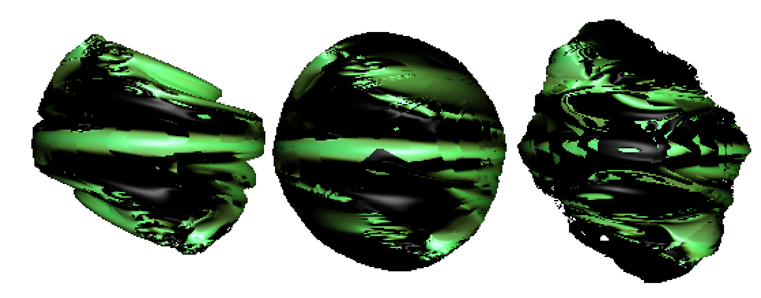}
{Early Experiments}
{Poorly defined distance parameters - exploring visual possibilities through random values.}
{1.0}

\bibliographystyle{acmsiggraph}



\let\oldthebibliography=\thebibliography
\let\endoldthebibliography=\endthebibliography

\renewenvironment{thebibliography}[1]{%
	\begin{oldthebibliography}{#1}%
		\setlength{\parskip}{0ex}%
		\setlength{\itemsep}{0.5ex}%
	}%
	{%
	\end{oldthebibliography}%
}


\bibliography{paper}


\figuremacroW
{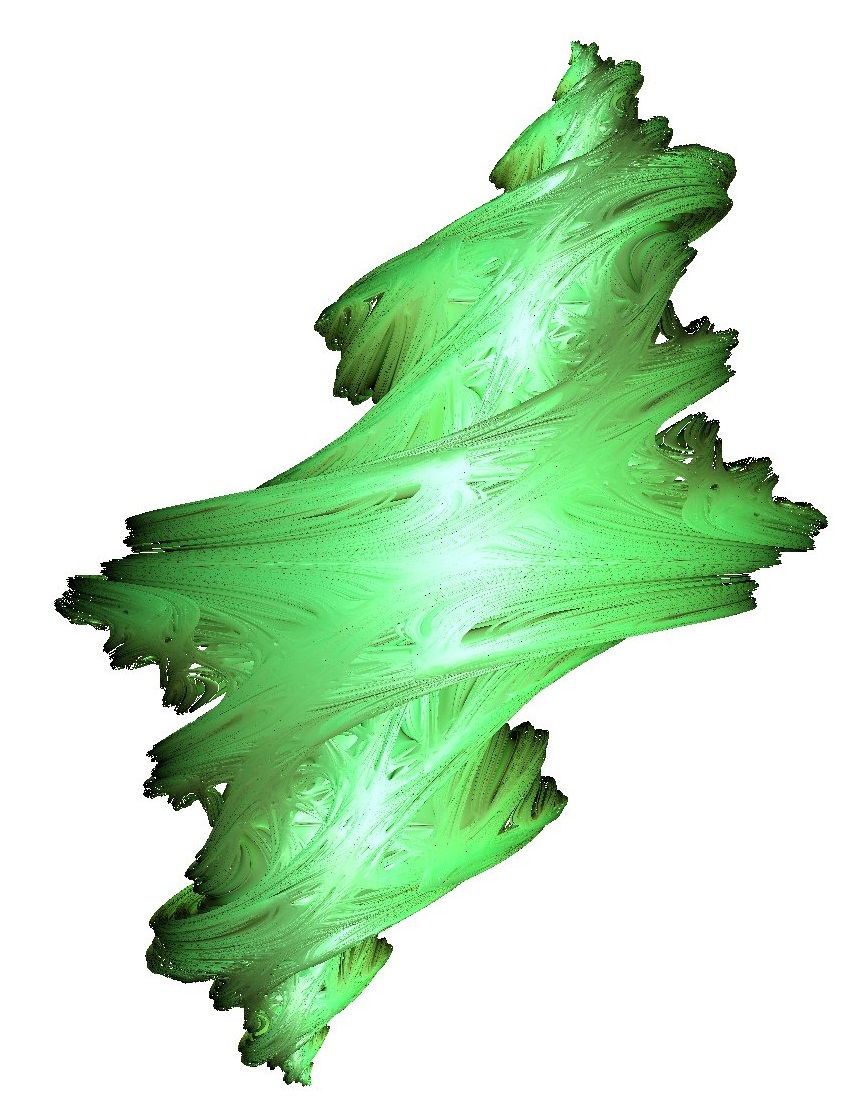}
{High-Detail}
{Emphasising detail within the dual-quaternion Julia set geometry by using a larger number of iterations. c = (-0.39054,-0.58679,0,0)(0.0,0.5632,0,0.05), n=15.}
{0.8}

\appendix

\section{Appendix} \label{sec:appendix}

\paragraph{Quaternion Operations}

The quaternions were discovered by Hamilton in 1843 as a method of performing 3-D multiplication \cite{kenwright2012beginners}. 
A quaternion $q$ is given by Equation \ref{eq:quatdef}.
Since we are combining quaternions with dual number theory, we give the elementary quaternion arithmetic operations.

\begin{equation}
q = [s,\vec{v}], (q_w,q_x,q_y,q_z)
\label{eq:quatdef}
\end{equation}

where $s$ scalar part is $s=q_w$ and vector part is $\vec{v}=(q_x,q_y,q_z)$. The four-tuple of independent real values assigned to one real axis and three orthonormal imaginary axes: $i,j,k$.

\begin{itemize}

\item \textbf{addition}: $q_1 + q_2 = [s_1,\vec{v}_1]+[s_2,\vec{v}_2] = [s_1+s_2,\vec{v}_1+\vec{v}_2]$

\item \textbf{additive identity}: $0 = [0,0]$

\item \textbf{scalar multiplication}: $kq=[ks,k\vec{v}]$

\item \textbf{multiplication}: $q_1 q_2 = [s_1,\vec{v}_1][s_2,\vec{v}_2] = [s_1s_2 - \vec{v}_1 \cdot \vec{v}_2, s_1\vec{v}_2 + s_2\vec{v}_1 + \vec{v}_1 \times \vec{v}_2]$

\item \textbf{multiplication identity}: $1 = [1,0]$

\item \textbf{dot product}: $q_1 \cdot q_2 = ( q_{1x}q_{2x} + q_{1y}q_{2y} + q_{1z}q_{2z} + q_{1w}q_{2w} )$

\item \textbf{magnitude}: $||q|| = \sqrt( s^2 + ||\vec{v}||^2 )$

\item \textbf{conjugate} $q^* = [s,-\vec{v}]$

\end{itemize}

\paragraph{Dual-Quaternion Operations}

The elementary arithmetic operations necessary for us to use dual-quaternions.

\begin{itemize}

\item \textbf{dual-quaternion}: $\zeta = q_r + q_d \varepsilon$

\item \textbf{scalar multiplication}:  $s \zeta = s q_r + s q_d \varepsilon$

\item \textbf{addition}: $\zeta_1 + \zeta_2 = q_{r1} + q_{r1} + (q_{d1}+q_{d2}) \varepsilon$

\item \textbf{multiplication}: $\zeta_1 \zeta_2 = q_{r1} + q_{r2} + (q_{r1}q_{d2} + q_{d1}q_{r2}) \varepsilon$

\item \textbf{conjugate}:  $\zeta^* = q_r^* + q_d^* \varepsilon$

\item \textbf{magnitude}: $||\zeta|| = \zeta \zeta*$

\end{itemize}
\noindent where $q_r$ and $q_d$ indicate the real and dual part of a dual-quaternion.

\end{document}